# New Distances to Four Supernova Remnants

S. Ranasinghe[1], D. A. Leahy[1], and Wenwu Tian[1,2]

*Department of Physics & Astronomy, University of Calgary, Calgary, Alberta T2N 1N4, Canada*

**Abstract:** Distances are found for four supernova remnants without previous distance measurements. HI spectra and HI channel maps are used to determine the maximum velocity of H I absorption for the four supernova remnants (SNRs). We examined $^{13}$CO emission spectra and channel maps to look for possible molecular gas associated with each SNR, but did not find any. The resulting distances for the SNRs are 3.5 ± 0.2 kpc (G24.7+0.6), 4.7 ± 0.3 kpc (G29.6+0.1), 4.1 ± 0.5 kpc (G41.5+0.4) and 4.5 ± 0 .4 - 9.0 ± 0.4 kpc (G57.2+0.8).

## 1. INTRODUCTION

Supernova remnants (SNRs) have a major impact on the state of the interstellar medium (ISM) of a galaxy (e.g. for a review see Cox [1]). They provide the thermal and kinetic energy that determines the structure of the ISM, and can drive outflows from the galaxy and trigger star formation. To determine the effects of SNRs, it is necessary to determine their occurrence rate, the energy of explosion and the density of the ISM where the explosion occurs.

Determining the distance to a SNR is an important first step in determining its size, age, explosion energy and evolutionary state. Utilizing H I absorption spectra is one method of obtaining the distance to a SNR. For the construction of H I absorption spectra, we use the method presented by Leahy and Tian [2]. However, in order to differentiate between false absorption features and real ones, a thorough investigation of individual H I channel maps is essential. To determine the distance and resolve kinematic distance ambiguity, we use the step-by-step process described by Ranasinghe and Leahy [3].

Using this method, we present new distances to four SNRs that do not have previous good measurements of distances: G24.7+0.6, G29.6+0.1, G41.5+0.4 and G57.2+0.8. The SNR G24.7+0.6 has been linked to the luminous blue variable star G24.73+0.69 [4]. Acero et al. [5] claimed that there is a possible X-ray detection consistent with the radio size. The SNR G29.6+0.1 is likely associated with the X-ray pulsar AX J1845-0258 [6]. However, there is very little known about the SNRs G41.5+0.4 and G57.2+0.8. For these four SNRs we also search for any morphological association with molecular clouds.

A brief introduction to the data and software used and the method of construction of H I absorption spectra is presented in section 2. The results for the four SNRs are given in Section 3. In Sections 4 and 5 we present the discussion and summary.

[1] Address correspondence to these authors at the department of Physics & Astronomy, 834 Campus Place N.W., University of Calgary, 2500 University Drive NW, Calgary, AB, Canada, T2N 1N4; Tel (403) 220-7192; E-mail: syranasi@ucalgary.ca, leahy@ucalgary.ca, wtian@ucalgary.ca

[2] Address correspondence to this author at the National Astronomical Observatories, CAS, Beijing 100012, China; Tel +86 010-64839807; E-mail: tww@bao.ac.cn

## 2. DATA ANALYSIS

### 2.1. Data and Software

We obtained the 1420 MHz continuum and H I -line emission data from the VLA (Very Large Array) Galactic Plane Survey (VGPS) [7]. To construct HI absorption spectra, we used MEANLEV, a software program in the DRAO EXPORT package. An advantage of MEANLEV is that one can extract 'on' (source) and 'off' (background) spectra defined by user-specified threshold continuum brightness ($T_B$) levels. One can first specify a spatial region, normally a box selecting a set of map pixels. Then source and background spectra are extracted using pixels above (for source) and below (for background) the given $T_B$ level. This maximizes the contrast difference in $T_B$, which maximizes the signal-to-noise in the H I spectrum.

The $^{13}$CO spectral line data was extracted from the Galactic Ring Survey of the Five College Radio Astronomical Observatory (FCRAO) 14 m telescope [8]. For the construction of the $^{13}$CO emission spectra, we used MEANLEV and the same source and background regions used for the H I spectra.

### 2.2. Construction of HI Absorption Spectra, CO Spectra and Kinematic Distances

The radiative transfer modeling follows from Leahy and Tian [2]. From the equation of radiative transfer, the H I absorption spectrum is

$$e^{-\tau_v} - 1 = \frac{T_{B,on}(v) - T_{B,off}(v)}{T_{B,on}^C - T_{B,off}^C}.$$

Here $T_B$ is the brightness temperature, the superscript 'C' denotes continuum and 'v' denotes velocity. The 'off' spectrum is the spectrum towards the background and the 'on' spectrum is the spectrum towards the source. The H I absorption spectrum of a continuum emission source (such as an SNR) is calculated. Radial velocities where there are likely absorption features are identified. It is essential that the H I channel maps are examined to see if the features in the spectrum are real, i.e. the spatial location of a decrease in H I emission is required to be coincident with and matched the shape of the continuum emission. Otherwise, we can see from the H I channel maps if the features are caused by excess H I emission in the source or background areas, and thus identify the features as false.

The $^{13}$CO spectra were extracted from the Galactic Ring Survey as noted above, and are plotted together with the H I absorption spectra below. We searched for $^{13}$CO emission features coincident with H I absorption features. If the $^{13}$CO emission feature is coincident with the highest velocity of real absorption with the SNR, there is a possibility of association of the $^{13}$CO with the SNR. Additionally, we examine the $^{13}$CO channel maps to see if the morphology of the $^{13}$CO emission matches the morphology of the SNR continuum emission. However for only one of the four SNRs studied here (G29.6+0.1), was there any indication of real association of $^{13}$CO. Once we obtain the radial velocity of the object (or upper and lower limits) from the H I spectra and channel maps, the distance (or upper and lower limits) to the source can be determined. For an object in circular orbit in the Galaxy at Galactocentric radius $R$, its radial velocity with respect to the local standard of rest (LSR) is

$$V_r = R_o \sin(l) \left( \frac{V(R)}{R} - \frac{V_o}{R_o} \right),$$

where $V(R)$ is the orbital velocity at $R$ [9]. Thus with an appropriate rotation curve the Galactocentric radius $R$ and the distance d can be obtained.

The distance has two solutions (near and far distances, called the kinematic distance ambiguity) for objects inside the solar circle, except at the tangent point. With H I absorption spectra, this kinematic distance ambiguity can be resolved (for a detailed description see Ranasinghe and Leahy [3]). If the absorption is not seen up to the tangent point, the object is at the near distance. If the absorption is seen up to the tangent point, the object is beyond the tangent point. For the first quadrant, any absorption seen at a negative velocity points to the object being located beyond the far side of the solar circle.

For the Galactic rotation curve we have adopted the Universal Rotation Curve (URC) presented by Persic et al. [10] and the parameters presented by Reid et al. [11] (their Table 5). The parameters are Galactocentric radius $R_0 = 8.34 \pm 0.16$ kpc, the orbital velocity of the sun of $V_0 = 241 \pm 8$ km s$^{-1}$, $a = 1.5$, $R_{opt} = R_0 \cdot (0.90 \pm 0.006)$, $V(R_{opt}) = 241 \pm 8$ and $\beta = 0.72$ .

To determine the error in the distance we follow the method in Ranasinghe and Leahy [3]. Due to the non-linearity of the equations, we find a set of distances for best fit parameters of $R_0$, $V_0$ and observed radial velocity $V_r$ and also for their lower and upper limits (a set of $3^3$ parameters and distances). The standard deviation of these values yields the error in distance. The errors in $R_0$ and $V_0$ are 0.16 kpc and 8 km s$^{-1}$ respectively [11]. We modeled the observed H I profile including a Gaussian velocity dispersion to obtain the tangent point velocity. The difference of the observed tangent point velocities and the URC gives an estimate for the peculiar motion of the gas of 4.7 km s$^{-1}$. Based on H I channel maps, the measurement error of the radial velocity $V_r$ is ± 2.4 km s$^{-1}$. The spectral resolution of the data is 1.5 km s$^{-1}$ [7]. The estimated peculiar motion of the H I and error in radial velocity added in quadrature yields a net error in $V_r$ of 5.3 km s$^{-1}$.

## 3. RESULTS

The four SNRs in this section are diffuse and relatively faint in 1420 MHz continuum (15 K to 20 K above local background). Because of the faint continuum, the extracted H I absorption spectra are noisy. To distinguish real absorption features from false ones requires a detailed investigation of the H I channel maps. The results presented here include H I maps for the most important absorption features relevant to distance determination.

### 3.1. G24.7+0.6

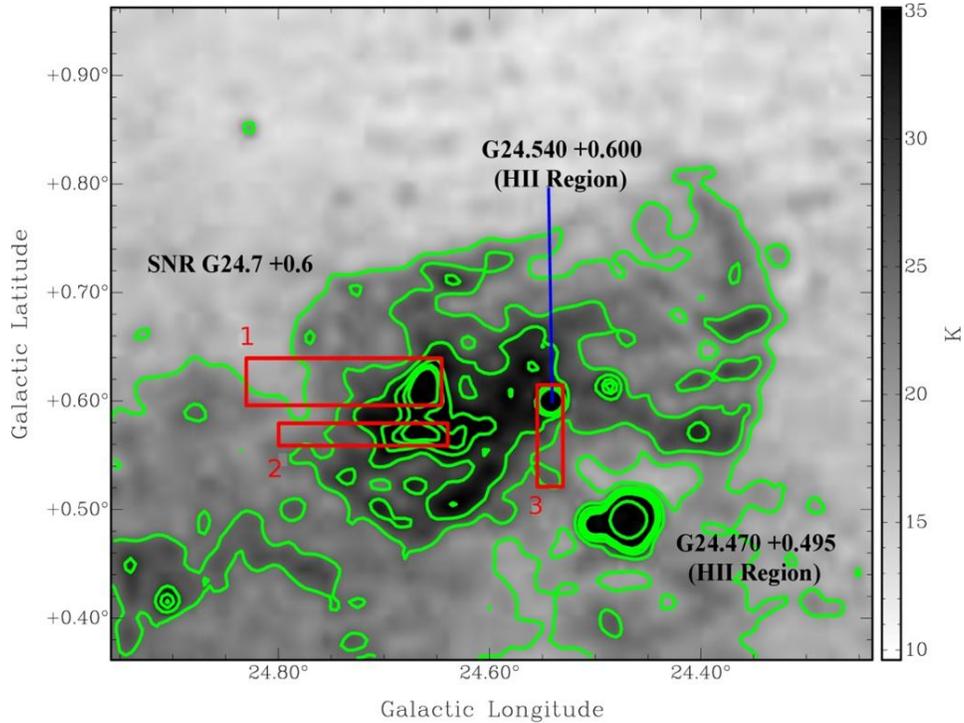

**Figure 1.** SNR G24.7 + 0.6 continuum image with contour levels (green) at 20, 25, 30, 35, 38, 40 and 100 K. The red box is the area used to extract H I and $^{13}$CO source and background spectra.

Figure 1 shows the 1420 MHz radio continuum image of G24.7+0.6 that we extracted from the VGPS survey. The spectrum extraction regions, from which both background and source spectra were found using MEANLEV, are shown by the red boxes in Figure 1. The brightest continuum regions of the SNR were included in Regions 1 and 2. Region 3 was chosen to extract the spectrum of the H II region G24.540+0.600 as a comparison for the SNR H I absorption spectra. The resulting H I spectra are shown in Figure 2. From the spectra of Regions 1 and 2 it appears that there may be absorption up to the tangent point.

The absorption features at 105 km s$^{-1}$ in the spectra are not verified in the H I channel maps (Figure 3 bottom panel). They are seen to be caused by excess H I emission in the background area. The absorption features seen at negative velocities for Region 1 and 2 spectra are similar to the tangent-point features. They are caused by excess H I in the background areas. The e$^{-\tau}$ > 1 features seen in the spectra are due to H I clouds in the chosen source region (Figure 3 top panel). From the H I images it is seen that 45 to 55 km s$^{-1}$ H I absorption features are real (Figure 3 middle panel). The H I channel maps indicate that the absorption features are not present beyond a velocity $V_r \geq 54.60$ km s$^{-1}$ which yields an estimated distance to the SNR as 3.5 kpc. The H II region G24.540+0.600 spectrum (Figure 3 bottom panel) and distance is discussed in Section 4.1.

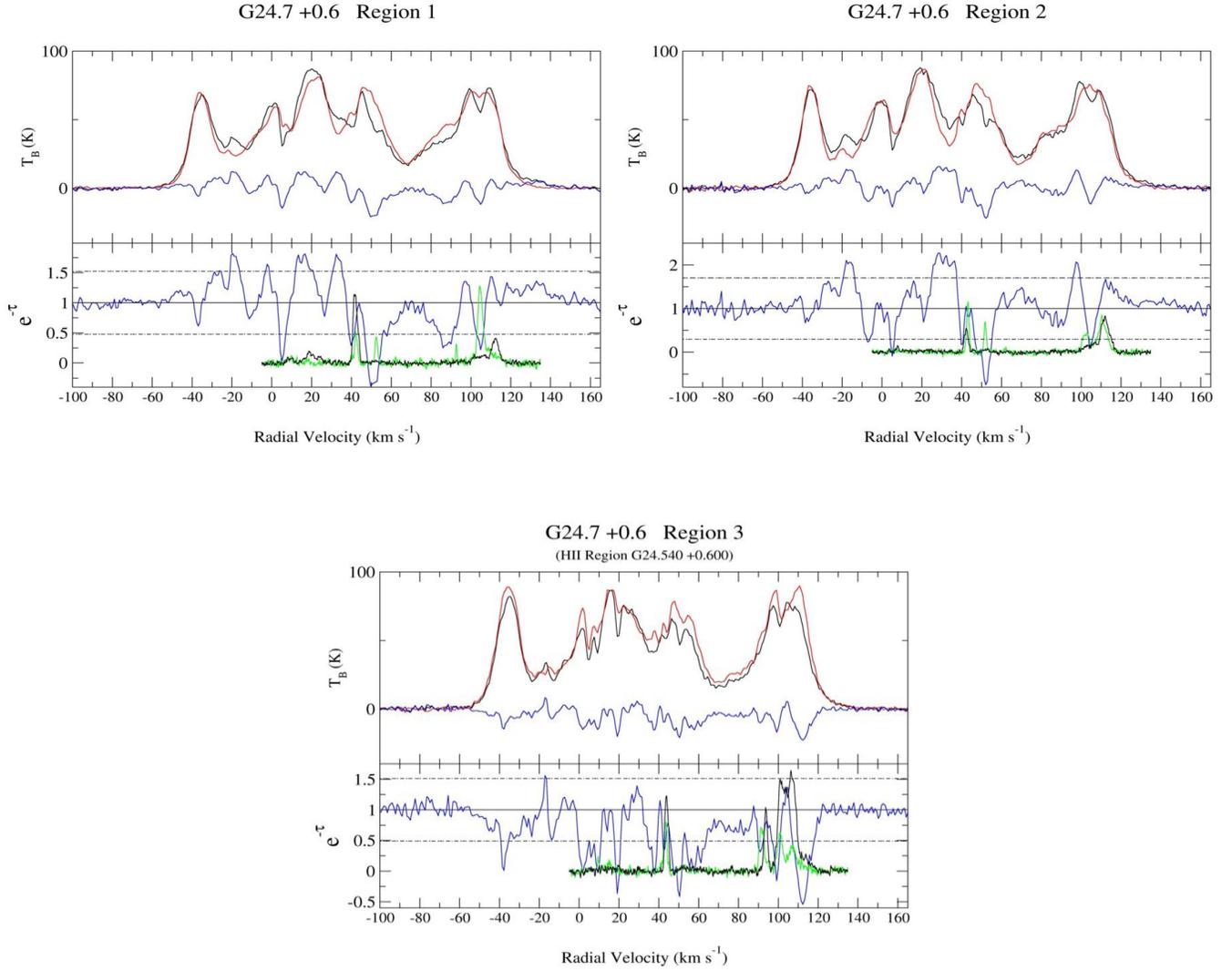

**Figure 2.** G24.7 + 0.6 spectra: The top panels show H I emission spectra: source spectrum (black), background spectrum (red) and difference (blue). The bottom panel gives the H I absorption spectrum (blue), the $^{13}$CO source spectrum (green) and the $^{13}$CO background spectrum (black). The dashed line is the $\pm 2\sigma$ noise level of the H I absorption spectrum. The URC tangent point velocity is +116.6 km s$^{-1}$ and distance is 7.6 kpc.

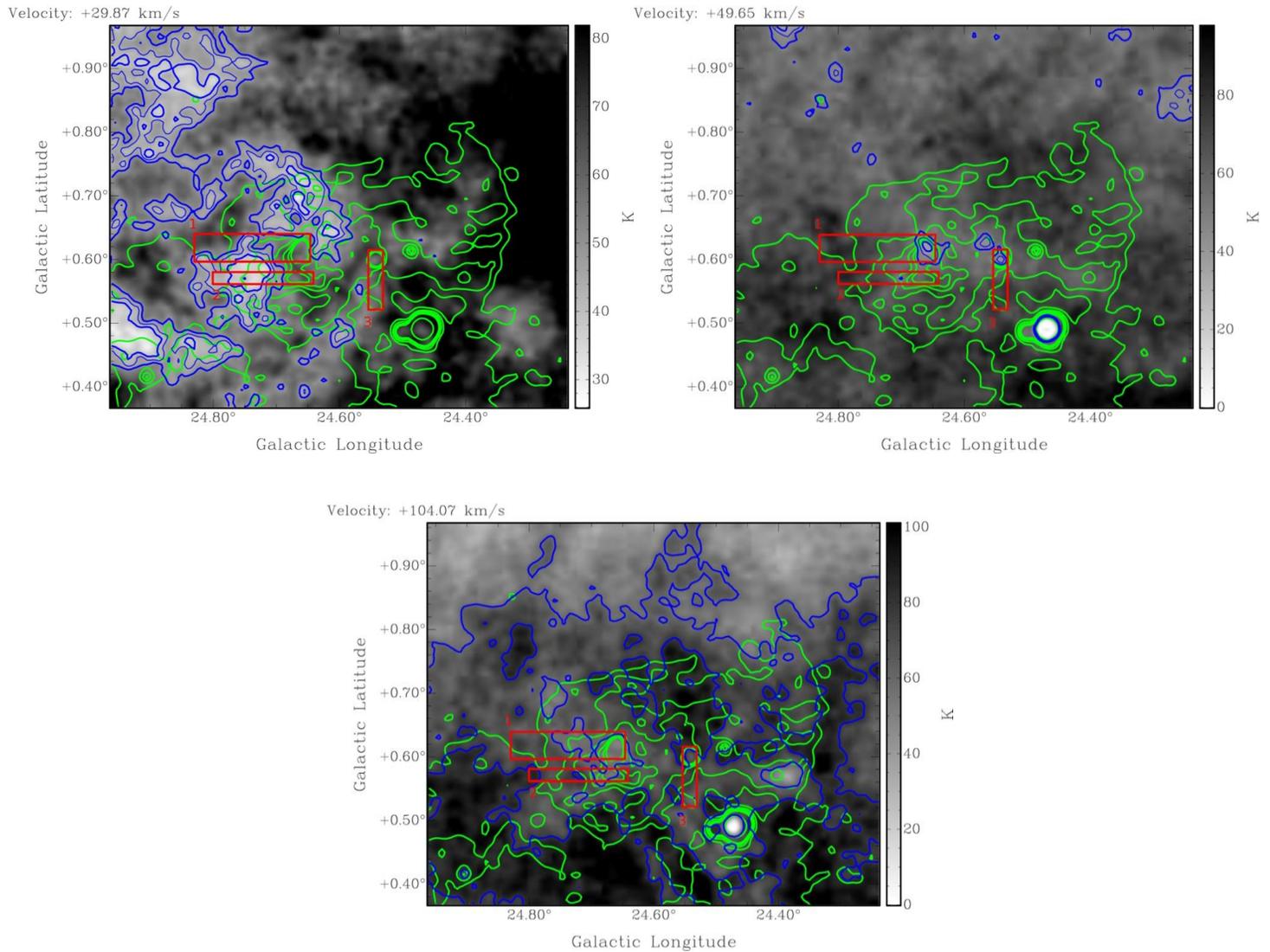

**Figure 3.** G24.7 + 0.6 H I channel maps +29.87, +49.65 and +104.07 km s⁻¹. The middle panel shows the H I intensity decreasing coincident with the bright continuum region of the SNR, indicating real absorption. The other two panels do not show such evidence for real absorption. The H I contour levels (blue) are at 40, 45 and 50 K for the channel maps +29.87 and +49.65 km s⁻¹ and 60 and 80 K for the channel map +104.07 km s⁻¹. The continuum contour levels (green) are at 20, 25, 30, 35, 38, 40 and 100 K.

## 3.2. G29.6 + 0.1

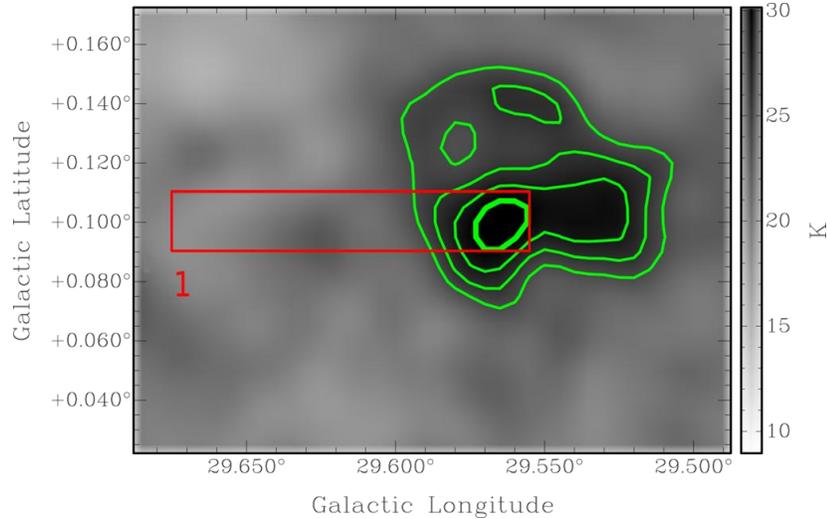

**Figure 4.** SNR G29.6 + 0.1 continuum image with contour levels (green) at 24, 26, 28 and 30 K. The red box is the areas used to extract H I and $^{13}$CO source and background spectra.

Figure 4 shows the 1420 MHz radio continuum image of G29.6+0.1. The brightest region of the SNR, with 1420 MHz brightness temperature 30 K, was used included in the source region (see red box in Figure 4 for location of area used for source and background). The resulting H I absorption spectrum is shown in Figure 5 and can be seen to have a high noise level. Thus we examine the H I channel maps to look for real absorption features. Three of the channel maps are shown here for illustration. The feature in the absorption spectrum at -25 km s$^{-1}$ (top panel of Figure 6) shows no correlation of decreased H I intensity with the continuum brightness of the SNR. Thus that feature is not real, but rather caused by the excess H I in the background region. Between the velocities of 100 and 110 km s$^{-1}$ a feature is present in the spectrum. However this feature is false, caused by excess H I in the background region (see channel map at 104 km s$^{-1}$ shown in Figure 6 bottom panel).

The maximum velocity where absorption occurs is at 80 km s$^{-1}$. The H I channel map at 78.5 km s$^{-1}$ is shown in the middle panel of Figure 6. The decrease in H I intensity is coincident with the maximum continuum intensity, consistent with real absorption. The lack of any evidence of absorption up to the tangent point, points to the near kinematic distance to the SNR. This places the SNR at a distance of 4.7 kpc.

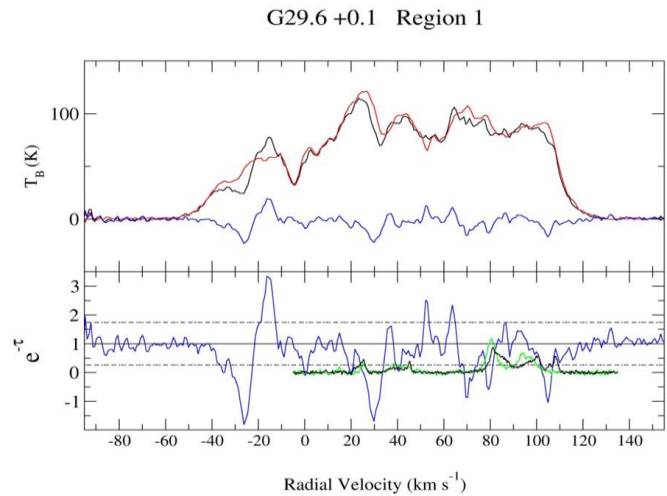

**Figure 5.** G29.6 + 0.1 spectra: The top panels show H I emission spectra: source spectrum (black), background spectrum (red) and difference (blue). The bottom panel gives the H I absorption spectrum (blue), the $^{13}$CO source spectrum (green) and the $^{13}$CO background spectrum (black). The dashed line is the ±2σ noise level of the H I absorption spectrum. The URC tangent point velocity is +108.1 km s$^{-1}$ and distance is 7.2 kpc.

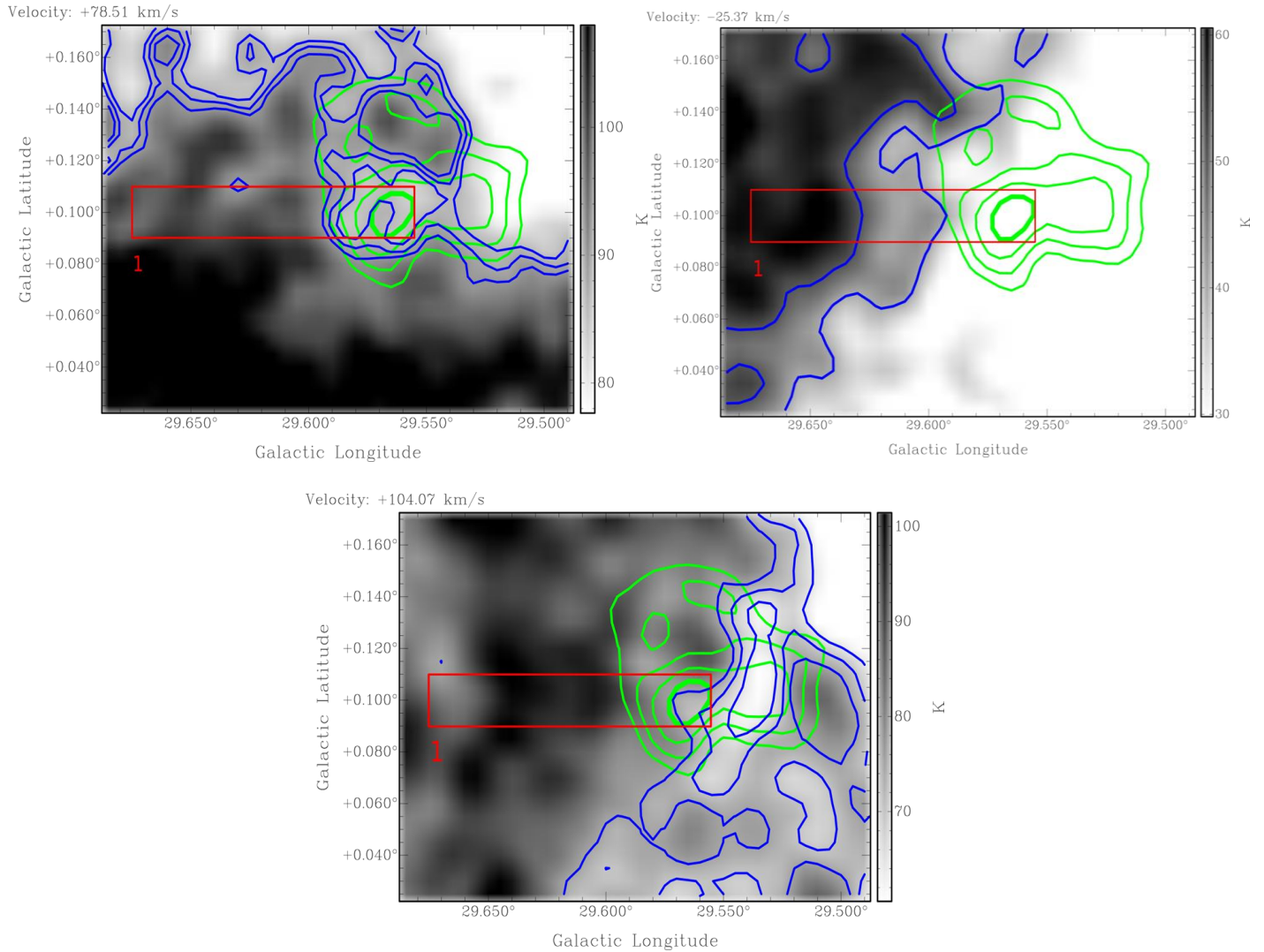

**Figure 6.** G29.6 + 0.1 H I channel maps -25.37, +78.51 and +104.07 km s⁻¹. The middle panel shows the H I intensity decreasing coincident with the bright continuum region of the SNR, indicating real absorption. The other two panels do not show such evidence for real absorption. The H I contour levels (blue) are at 54 and 58 K for the channel map -25.37 km s⁻¹, 86, 88 and 90 K for the channel map +78.51 km s⁻¹ and 65, 70 and 75 K for the channel map +104.07 km s⁻¹. The continuum contour levels (green) are at 24, 26, 28 and 30 K.

### 3.3 G41.5 + 0.4

Figure 7 shows the 1420 MHz radio continuum image of G41.5+0.4. The three brightest regions of the SNR G41.5 + 0.4 were chosen for making H I spectra, which are shown in Figure 8. All three spectra are noisy and show few consistent features. The features with $e^{-\tau} > 1$ or with $e^{-\tau} < 0$ were verified to be excess H I emission in either the source region or in the background region, and not caused by absorption.

Figure 9 shows selected H I channel maps. The upper left panel shows the map for -27.85 km s⁻¹, illustrating the excess H I emission in the background area of regions 1 and 2. The upper right panel shows the map for +63.67 km s⁻¹, illustrating the correspondence between the decrease in H I intensity and the brightest continuum emission regions of the SNR. This indicates real absorption. The lower left panel shows the map for +76.86 km s⁻¹. This illustrates a lack of correspondence between the decrease in H I intensity and the brightest continuum emission regions of the SNR. This indicates no absorption at this velocity and shows that the $e^{-\tau} > 1$ in the region 3 H I spectrum (in Figure 8) is caused by excess emission in the source area of the region. The lower right panel shows the map for +83.46 km s⁻¹. There is no correspondence between the decrease in H I intensity and the brightest continuum emission regions of the SNR, thus no real

H I absorption at this velocity.

In summary, from examination of each H I channel map, we find no evidence of absorption up to the tangent point at +85 km s⁻¹. There is no absorption present in the negative velocity range. There is clear absorption up to +64 km s⁻¹ which is consistent in the H I channel maps. Because there is no absorption present up to the tangent point or at the negative velocities, this places the SNR at the near kinematic distance. The SNR at a distance for a radial velocity of +64 km s⁻¹ is 4.1 kpc.

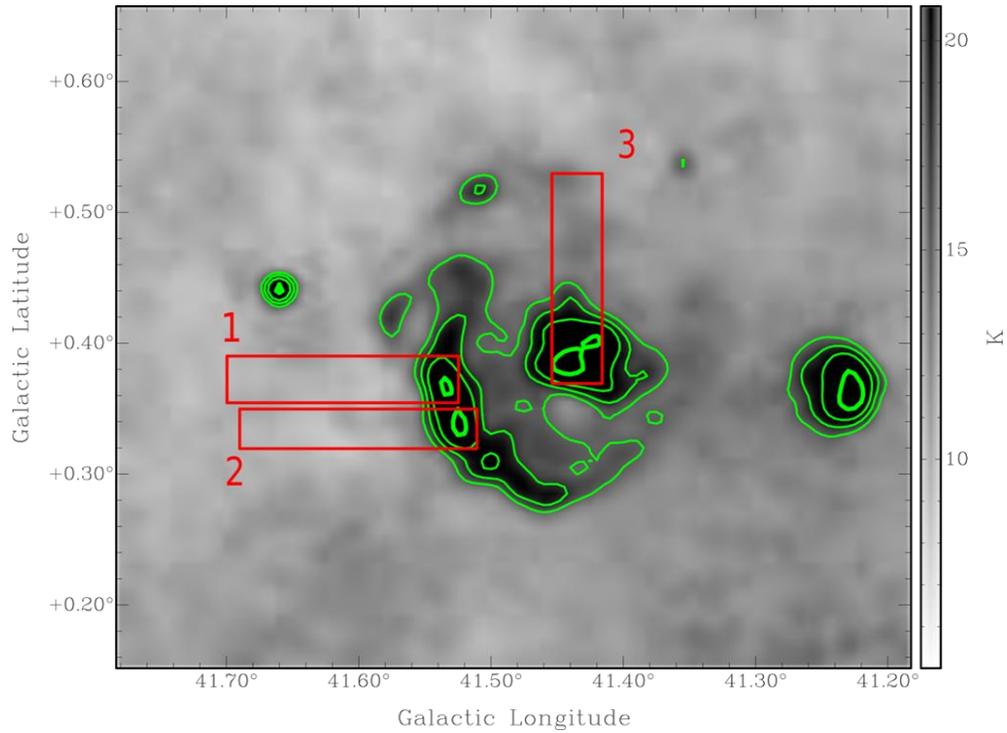

**Figure 7.** SNR G41.5 + 0.4 continuum image with contour levels (green) at 15, 18, 22 and 30 K. The red boxes are the areas used to extract H I and ¹³CO source and background spectra.

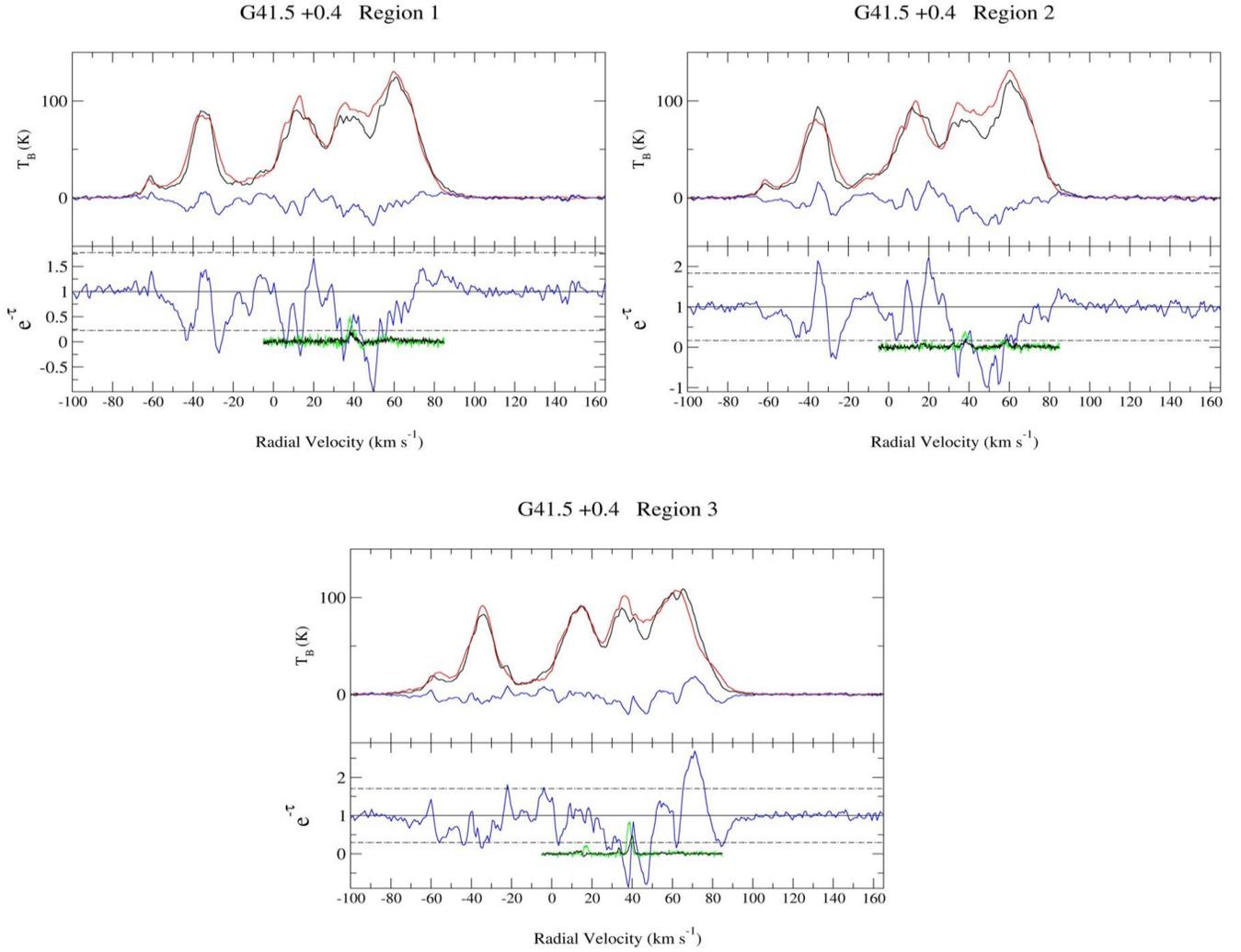

**Figure 8.** G41.5 + 0.4 spectra: The top panels show H I emission spectra: source spectrum (black), background spectrum (red) and difference (blue). The bottom panel gives the H I absorption spectrum (blue), the $^{13}$CO source spectrum (green) and the $^{13}$CO background spectrum (black). The dashed line is the $\pm 2\sigma$ noise level of the H I absorption spectrum. The URC tangent point velocity is +78.4 km s$^{-1}$ and distance is 6.2 kpc.

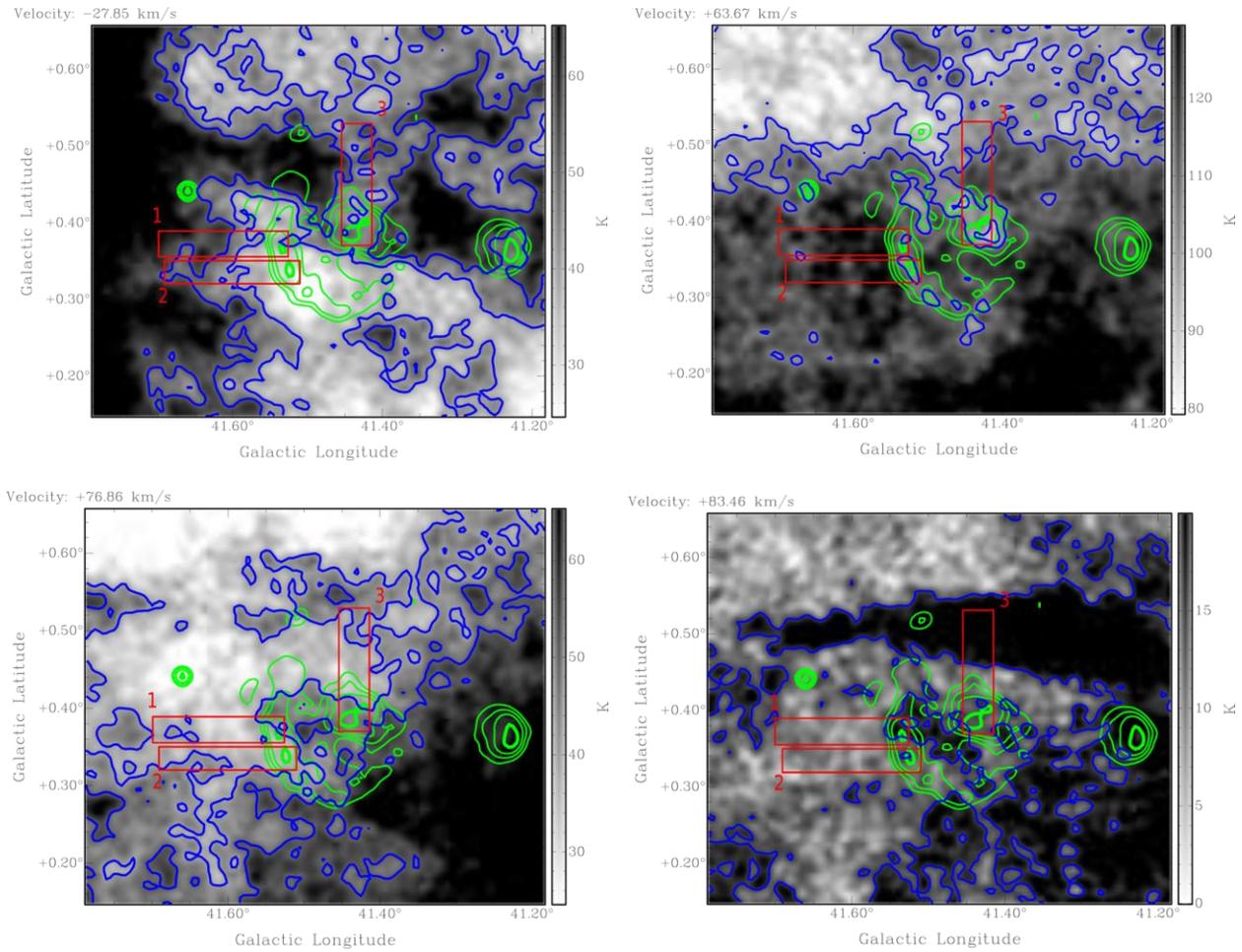

**Figure 9.** G41.5+0.4 H I channel maps -27.85, +63.67, +78.86 and +83.46 km s⁻¹. The top right panel shows the H I intensity decreasing coincident with the bright continuum region of the SNR, indicating real absorption. The three other panels do not show such evidence for real absorption. The H I contour levels (blue) are at 40 and 50 K for the channel map -27.85 km s⁻¹, 95 and 105 K for the channel map +63.67 km s⁻¹, 40 and 50 K for the channel map +78.86 km s⁻¹ and 15 K for the channel map +83.46 km s⁻¹. The continuum contour levels (green) are at 15, 18, 22 and 30 K.

### 3.4. G57.2+0.8

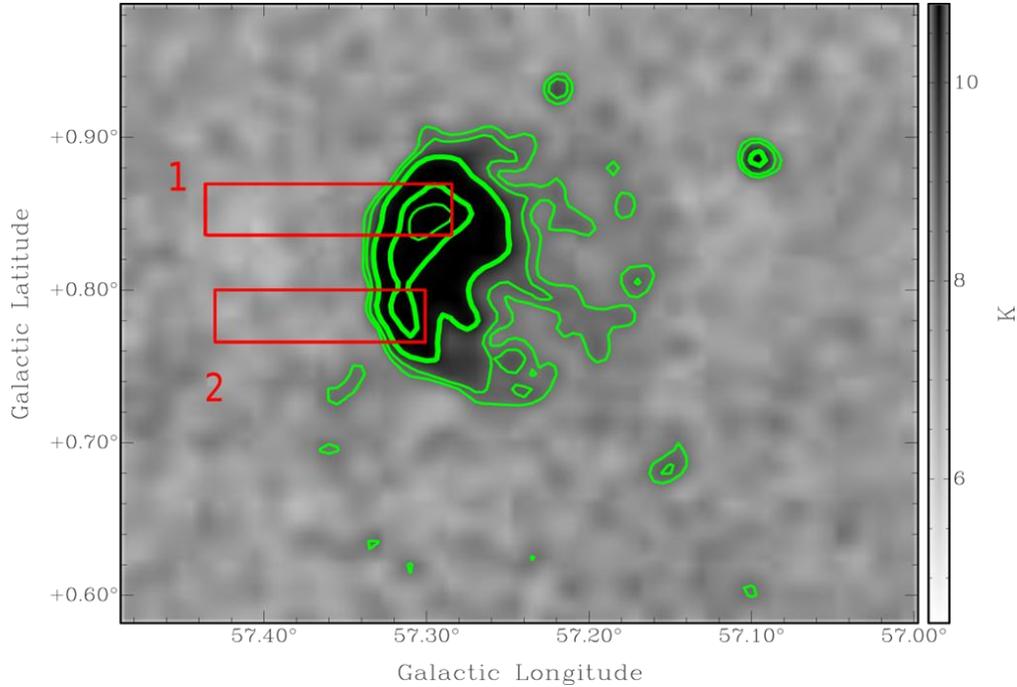

**Figure 10.** SNR G57.2+0.8 continuum image with the contour levels (green) are at 8, 8.5, 10, 15 and 18 K. The red boxes are the areas used to extract H I and $^{13}$CO source and background spectra.

The continuum image of G57.2+0.8 and the regions for extraction of H I absorption spectra are shown in Figure 10. The spectra are shown in Figure 11. The $^{13}$CO survey covers a longitude range of 18 to 55.7 and a latitude range of -1 to +1 [8]. This SNR is located beyond the longitude range covered by the survey and therefore does not include the $^{13}$CO emission spectra. The SNR is relatively faint yielding noisy H I spectra. Region 1 includes the brightest region of the SNR, but both spectra are included in the analysis in order to compare absorption features.

The top panel of Figure 12 shows the H I channel map at -46 km s$^{-1}$ and shows no correspondence between decreased HI and increased continuum intensity. This shows that the features in the H I spectrum near -46 km s$^{-1}$ are not real absorption. No real absorption was found at negative velocities, giving an upper limit of distance as the far side of the solar circle at 9.0 kpc. It is seen from the spectra that there are absorption features up to the tangent point at 45 km s$^{-1}$. This absorption up to the tangent point is verified in the individual H I channel maps. The bottom panel of Figure 12 shows the map at +44.7 km s$^{-1}$. This shows a good correspondence between decreased H I and increased continuum intensity, thus real absorption. Therefore the lower limit distance to the SNR is the tangent point distance of 4.5 kpc. We place the SNR between 4.5 and 9.0 kpc.

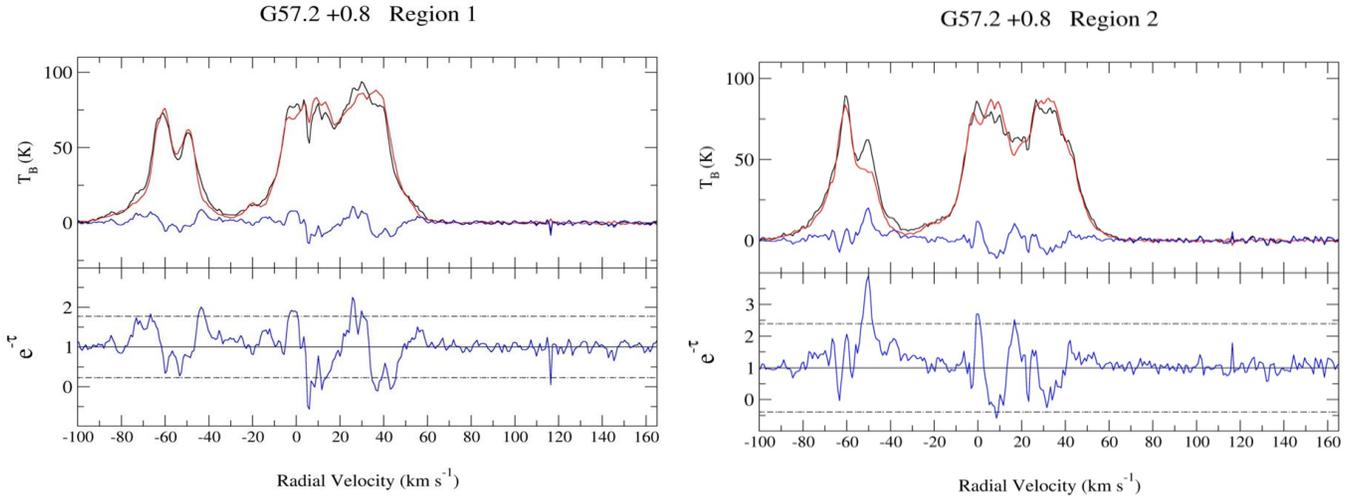

**Figure 11.** G57.2+0.8 spectra: The top panels show H I emission spectra: source spectrum (black), background spectrum (red) and difference (blue). The bottom panel gives the H I absorption spectrum (blue), the $^{13}$CO source spectrum (green) and the $^{13}$CO background spectrum (black). The dashed line is the ±2σ noise level of the H I absorption spectrum. The URC tangent point velocity is +38.3 km s$^{-1}$ and distance is 4.5 kpc.

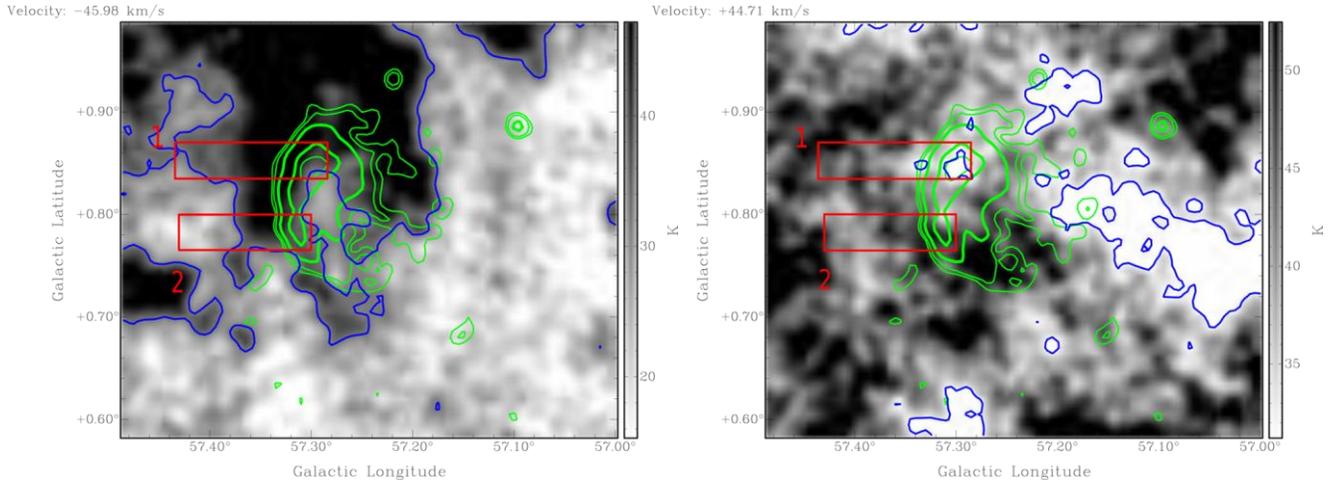

**Figure 12.** G57.2+0.8 H I channel maps -45.98 and +44.71 km s$^{-1}$. The bottom panel shows the H I intensity decreasing coincident with the bright continuum region of the SNR, indicating real absorption. The top panel does not show such evidence for real absorption. The H I contour levels (blue) are at 32 K. The continuum contour levels (green) are at 8, 8.5, 10, 15 and 18 K.

## 4. DISCUSSION

### 4.1. G24.7+0.6

The SNR G24.7+0.6 H I absorption distance is 3.5 kpc. G24.7+0.6 is a SNR 30′×15′ in size, with a filled center and a faint shell. Becker and Helfand [12] claimed a compact H II region 0.7″ in size superimposed on the SNR at l = 24.677 and b = +0.5495. This does not affect our conclusion on distance of SNR G24.7+0.6 because the same maximum velocity of real absorption is seen for both regions of the SNR, the lower one including the above H II region and the upper one not including it.

The SNR is located near the H II regions G24.540+0.600 and G24.470+0.495. Using the International Galactic Plane Survey, Jones and Dickey [13] inferred that the H II regions G24.540+0.600 and G24.470+0.495 are at kinematic distances of 19.43$^{+1.93}_{-1.96}$ kpc and 7.74$^{+1.45}_{-0.88}$ kpc respectively. Neither of these are consistent with the H I absorption distance of 3.5 kpc, so are not associated with the SNR. We studied the HI absorption spectrum of the H II region G24.540+0.600 (Figure 2 bottom panel). It shows absorption up to the tangent point (116 km s$^{-1}$). We verified this using the channel maps, e.g. absorption seen in the 104.07 km s$^{-1}$ (bottom panel of Figure 3). The maximum velocity of absorption is -38.6 km s$^{-1}$ which gives a distance of 20.6 kpc using the Reid et al. [11] rotation curve. This is consistent with the Jones and Dickey [13] distance. The H I spectrum and channel maps of H II region G24.470+0.495 show absorption up to the tangent point, but none at negative velocities. This H II region has $V_r = 29.87$ km s$^{-1}$ which places it at far side of the tangent point at 12.9 kpc. Thus G24.470+0.495 is considerably farther than the tangent point distance of 7.6 kpc.

Petriella et al. [4] studied the molecular environment of the luminous blue variable (LBV) star G24.73+0.69 that is located near the SNR. They placed it at 3.5 kpc adopting a systemic velocity of 42 km s$^{-1}$ of the molecular shell and suggested the progenitor of the SNR G24.7+0.6 and the LBV star both were formed from the same natal cloud. Using the systemic velocity of 42 km s$^{-1}$ from Petriella et al. [4] and the Reid et al. [11] rotation curve, the revised distance to the LBV star G24.73+0.69 is 2.9 kpc. The LBV star could be associated with the SNR if it has a peculiar velocity of +14 km s$^{-1}$. This would place the LBV star also at a distance of 3.5 kpc.

The error of the SNR distance is calculated using the method described in Section 2.2 and is 0.2 kpc. The size of the SNR is $30.5 \pm 1.7 \times 15.3 \pm 0.9$ pc of the major and minor axes respectively. These values are summarized in Table 1.

## 4.2. G29.6+0.1

Gaensler et al. [6] reported Very Large Array observations of the slow X-ray pulsar AX J1845-0258, which is physically associated with the SNR. The 5.1′ radio emission shell of the SNR is linearly polarized with a non-thermal spectral index. The SNR is thought to be more than 8000 years old. Vasisht et al. [14] determined the distance to the pulsar to be 5 - 15 kpc based on X-ray absorption measurements. The distance to the SNR was suggested as 10 kpc due to its association with AX J1845-0258. The chosen region for our analysis coincides with the bright spot of the Gaensler et al. [6] 5 GHz image (their Figure 1, left panel).

Kilpatrick et al. [15] detected $^{12}$CO emission toward the SNR at a velocity of +94 km s$^{-1}$ but gave no good evidence of an association (their Figure 12). The $^{13}$CO spectra from the source and background regions inside the red box (Figure 4) are shown in Figure 5. These show molecular clouds at 65, 79, 85 and 95 km s$^{-1}$. None of these molecular clouds morphologically match the boundary of the SNR. However the cloud at 79 km s$^{-1}$ matches the highest velocity of absorption of the SNR and thus could be associated with the SNR.

The distance to the SNR is $4.7 \pm 0.3$ kpc and the diameter is $6.8 \pm 0.4$ pc. The distance is consistent with the approximate lower limit distance presented by Vasisht et al. [14].

## 4.3. G41.5+0.4

Kaplan et al. [16] first presented an image of the SNR and noted its complex morphology, consisting of a brighter rim to the left and a compact core to the right suggestive of a PWN. The SNR is ~12′ in size. Alves et al. [17] confirmed the synchrotron nature of the SNR presenting free-free and synchrotron maps. There is no previous distance estimation for G41.5+0.4. We place G41.5+0.4 at a distance of $4.1 \pm 0.5$ kpc. The diameter of the SNR is 16.7 ± 2.0 pc.

## 4.4. G57.2 + 0.8

Also known also as 4C21.53, the 13′×10′ SNR G57.2+0.8 has a spectral index of 0.62 and consists of a non-thermal arc [18]. Kilpatrick et al. [15] used a Σ-D relation to estimate the distance to G57.2+0.8 as 8.2 kpc. Park et al. [19] used another Σ-D relation to estimate the distance as 14.3 kpc.

We place the SNR G57.2+0.8 between the tangent point distance of 4.5 kpc and the far-side of the solar circle of 9 kpc. We used the method of error calculation in distance described in Section 2.2. For the tangent point distance, many parameter combinations yield no solutions to the equations for $V_r(d)$ (where $V_r \geq V(R) - V_0 \sin(l)$). Thus the standard deviation of solutions for d is artificially low. To obtain a more reasonable estimate of the error at the tangent point, we constructed a model spectrum and compared the tangent point velocity with the URC rotation curve velocity. We find the corresponding error as 0.4 kpc for both lower and upper limit distances.

## 4.5. Evolutionary states

With distances now measured, we apply a basic Sedov model [20] to estimate the ages of the four SNRs. None of the four SNRs have X-ray observations. Thus we cannot use the X-ray emission to determine the local ISM density for the SN explosion, as done in [21] using the models of Leahy and Williams [22]. Leahy [21] found that the mean explosion energy for 50 SNRs in the Large Magellanic Cloud was $5\times10^{50}$ erg, so we use that value. The range in local ISM density for the LMC was found to be larger (from ~$10^{-3}$ cm$^{-3}$ to ~10 cm$^{-3}$) than the range of explosion energy. Because the SNRs we are considering are inside the solar circle, in a higher density part of the Milky Way, we use 1 cm$^{-3}$ as for the nominal ISM density.

SNR nominal ages were found by applying a Sedov model with explosion energy $E_0 = 5 \times 10^{50}$ erg, local ISM density $n_0 = 1$ cm$^{-3}$ and our new distances. The resulting age for G24.7+0.6 is 9100 yr, for G29.6+0.1 is 440 yr and for G41.5+0.4 is 4100 yr. G57.2+0.8 has a Sedov age between 3900 yr (for distance 4.5 kpc) and 22,000 yr (for distance 9 kpc). Not having a measurement of the local ISM density creates a larger uncertainty in the model SNR age than any other factor. E.g., a difference in local ISM density by a factor of 10 results in an age difference by a factor of 3.16. The Sedov age estimate for G29.6+0.1 is quite low. As discussed in Section 4.2, this SNR may be associated with a molecular cloud, indicating a higher density. If the ISM density was 10 to 100 cm$^{-3}$, quite feasible if it is associated with a molecular cloud, then the Sedov age is 1400 to 4400 yr. Thus for G29.6+0.1 we use a range of $n_0 = 1$ - 100 cm$^{-3}$.

**Table 1. Distances to supernova remnants.**

| Source | Literature Dist [a] (kpc) | Ref. | $V_r$ (km s$^{-1}$) | KDAR [b] | New Dist (kpc) | Angular size [c] (arcmin) | Size [c] (pc) | Sedov Age [d] (kyr) |
|---|---|---|---|---|---|---|---|---|
| G24.7+0.6 | - | - | 54.6 | N | $3.5 \pm 0.2$ | 30 15 | $30.5 \pm 1.7$  $15.3 \pm 0.9$ | 9.1 |
| G29.6+0.1 | $10 \pm 5$ [X] | 14 | 80.16 | N | $4.7 \pm 0.3$ | 5  5 | $6.8 \pm 0.4$  $6.8 \pm 0.4$ | $0.44 - 4.4$ |
| G41.5+0.4 | - | - | 63.67 | N | $4.1 \pm 0.5$ | 14 14 | $16.7 \pm 2.0$  $16.7 \pm 2.0$ | 4.1 |
| G57.2+0.8 | 8.2 [S]  14.3 [S] | 15 19 | $V_{TP} - 0$ | $V_{TP} - $ F | $4.5 \pm 0.4 - 9.0 \pm 0.4$ | 13 12 | $25.5 \pm 10$  $23.6 \pm 9.3$ | $3.9 - 22$ |

Notes:
[a] Distance method - x: X-ray absorption, s: $\Sigma$-D relation.
[b] KDAR- Kinematic Distance Ambiguity Resolution, indicating near (N), far (F): upper limit solar circle or tangent point (TP) distance.
[c] Major axis $\times$ minor axis in radio continuum.
[d] Sedov age with $E_0 = 5 \times 10^{50}$ erg, $n_0 = 1$ cm$^{-3}$, except for G29.6+0.1 (see text) with $n_0 = 1 - 100$ cm$^{-3}$.

## 5. SUMMARY

We have used H I absorption spectra and H I channel maps to find the maximum velocity of absorption for four SNRs which do not have previous distance measurements. The H I channel maps were used to distinguish between real and false features in the H I absorption spectrum. The resulting distances and distance limits for the four SNRs are given in Table 1. The angular sizes (major and minor axes) measured from the 1420 MHz radio continuum images are given in Table 1. The resulting physical sizes, using the new distances are also given. We have applied Sedov models to estimate the ages of the four SNRs, and find ages typical of SNRs in the Sedov phase.

### ACKNOWLEDGEMENTS

This work was supported in part by a grant from the Natural Sciences and Engineering Research Council of Canada. We would also like to thank the referee for the insightful comments and important suggestions that have improved this work.